\documentclass[aps,onecolumn,groupedaddress,longbibliography]{revtex4-1}
\usepackage{graphicx}
\usepackage[T1]{fontenc}
\usepackage{bm}
\usepackage{mathptmx}
\DeclareMathAlphabet{\mathcal}{OMS}{cmsy}{m}{n}
\DeclareSymbolFont{largesymbols}{OMX}{cmex}{m}{n}
\usepackage{amsmath}
\usepackage{amssymb}
\usepackage{color}
\usepackage{titlesec}

\makeatletter
\renewcommand{\section}{\@startsection{section}{1}{0mm}{-\baselineskip}{0\baselineskip}{\bf\leftline}}
\makeatother

\begin{document}
	
	\preprint{}	
	\title{Vertical-supercooling-controlled interfacial instability for a spreading liquid film}
	
	\author{Li Chen}
    \author{Feng Wang}
    \author{Yingrui Wang}
    \author{Peng Huo}
    \author{Yuqi Li}
    \author{Xi Gu}
    \author{Man Hu}
    \author{Daosheng Deng}
    \email{dsdeng@fudan.edu.cn}
\affiliation{Department of Aeronautics and Astronautics, Fudan University,Shanghai, 200433, China} 

	
	\date{\today}
	
\begin{abstract}
Thermal effect is essential to regulate the interfacial instabilities for diverse technology applications. Here we report the fingering instability at the propagation front for a spreading liquid film subjected to the supercooling at the vertical direction. We find the onset timescale of hydrodynamic instability is strongly correlated with that of the vertical solidification process. This correlation is further validated in a non-uniform geometry, demonstrating the capability of controlling fingering instability by structure design. We attribute the identified interfacial instability to a pronounced thermo-viscous effect, since the rapidly increased viscosity of propagation front undergoing solidification can significantly enhance the mobility contrast locally in the vicinity of the spreading front, consequently producing the instability analogous to viscous fingering. This work offers another valuable dimension by gating the vertical temperature to exploit the interfacial stabilities and steer liquid flow, consequently shedding light on the microfluidic cooling for electronics, and the advanced functional fibers and fabrics.

\end{abstract}
	
	\maketitle
	
\section{Introduction}
Interfacial instabilities and the correlated intriguing patterns are not only ubiquitous in nature, such as the snowflakes and patterns of bacterial colonies \cite{langer1980instabilities, ben2000cooperative}, but also essential for technology applications, such as the coating processing, microfluidics, and Lab-on-a-Chip \cite{craster2009dynamics, stone2004engineering,darhuber2005principles}. Particularly, thermal effect plays an indispensable role for the interfacial instabilities \cite{davis1987thermocapillary,schatz2001experiments}. For example, during the unidirectional freezing, Mullins-Sekerka instability occurs at the moving planar liquid-solid interface \cite{mullins1964stability}; and under a unidirectional temperature gradient,  the thermocapillary instability appears \cite{boos1997thermocapillary}. At the microscopic scale, the heat transfer becomes even more pronounced for interfacial stabilities, significantly influencing the fluid dynamics and the subsequently produced patterns \cite{cazabat1990fingering,kataoka1999patterning}. In spite of these extensive studies, heat transfer and temperature gradient are mainly confined within the liquid film.


Recently, by surveying an impacted droplet on a cold substrate \cite{Soninphysfluid1997}, diverse patterns have been clearly revealed, such as the crack pattern changing from a 2D fragmentation to a hierarchical fracture \cite{ghabache2016frozen}, the self-peeling of impacting droplets \cite{RuiterNatphy2018}, and morphology evolution from conical tips to toroidal shapes \cite{hu2020frozen}. Fingering growth emerges in a  binary droplet,  which freezes from the outside prior to the impacting on the cold surface  \cite{kant2020pattern}. At a microscopic level, the nucleation process and the associated  solidification or crystallization  during the droplet spreading and cooling have been revealed to be responsible for the eventual morphology and the arrest of the contact line \cite{Pallav2020fastfreezing,Koldeweij2021Initial,Grivet2022Contact,LambleyNatphys2023}. However, the freezing of impacted droplets and the correlated spreading process are extremely complicated, because of the inherent complex interplay of the rapid impacting hydrodynamics, the non-uniform liquid thickness, the transient heat transfer, and the intricate solidification.

In this work, using a Hele-Shaw cell, a liquid film with uniform thickness is produced and subjected to cooling along the side walls, which not only circumvents the rapid dynamic deformation of the liquid film but also enables us to focus on the interfacial instability quantitatively. We observe fingering instability at the spreading front, and identify the onset of the instability is primarily determined by the solidification process at the vertical direction. Additionally, we demonstrate this vertical-supercooling-induced fingering instability can be further engineered simply through structure designing of a non-uniform geometry.  Moreover, the unstable wavelength increases with the critical velocity or the inlet flow rate, which can be explained by the thermal-viscous instability.  

	\begin{figure}[b]
		\centering
		\includegraphics[width=0.8\textwidth]{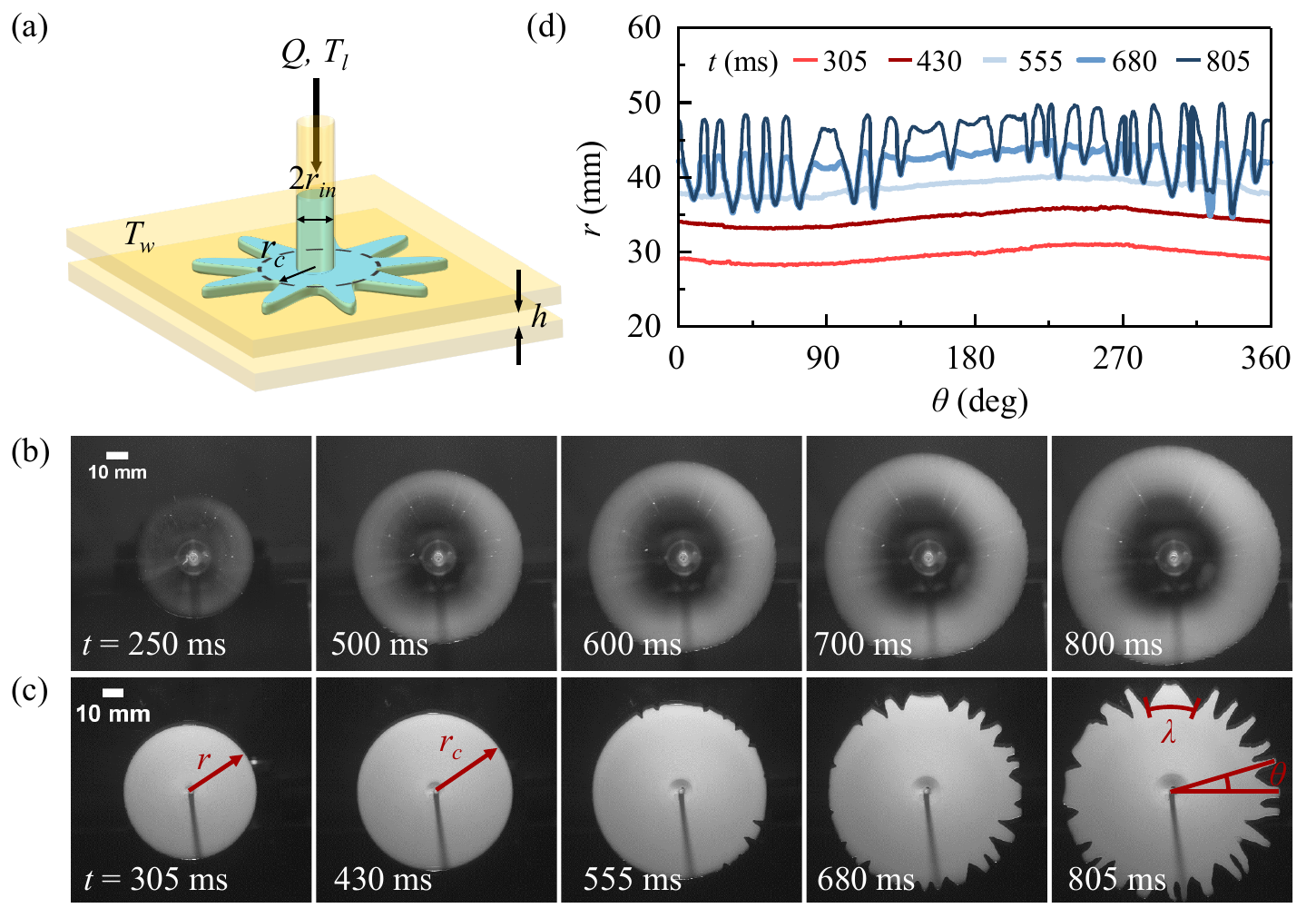}
		\caption{Spatiotemporal evolution of fingering instability for an injected liquid at an elevated temperature into a Hele-Shaw cell subjected to the side-wall supercooling. (a) Schematic diagram of the experimental setup; (b, c) Snapshot for a stable interface at a higher $T_w \approx T_m$ ($T_w =67$ $^{\circ}$C) and fingering instability at a lower $T_w<T_m$ ($T_w=25 \ ^{\circ}$C) for a paraffin wax film in a glass cell ($Q=4 $ ml/s,  $h = $ 400 $\mu$m); (d) Temporal evolution of the azimuthal undulations along the radius $ r(\theta) $ for the interface in row (c).}
		\label{fig:observedinstability}
	\end{figure}

\section{Fingering instability}  As shown in Fig.\ref{fig:observedinstability}a (Supplementary Note 1), paraffin wax is molten into a viscous state at an elevated temperature ($T_l = 95 \ ^{\circ}$C) above its melting point ($T_m = 69\ ^{\circ}$C). Subsequently, this viscous liquid is injected into a Hele-Shaw cell consisting of two glass plates with a gap thickness $h$ = 400 $\mu$m through a central inlet ($2r_{in}= 2.5$ mm). The flow rate of injection is varied by a syringe pump ($Q= 0.25-4$ ml/s), and the cooling process is controlled or gated by the wall temperature of Hele-Shaw cell ($T_w$). At a higher $T_w= 67\ ^{\circ}$C ($T_w \approx T_m$), when the injected liquid spreads within the Hele-Shaw cell, the moving front is stable, and the circular symmetry of the interface is preserved (Fig. \ref{fig:observedinstability}b).

\begin{figure}[htb]{}
    \center
     \includegraphics[width=0.8\textwidth]{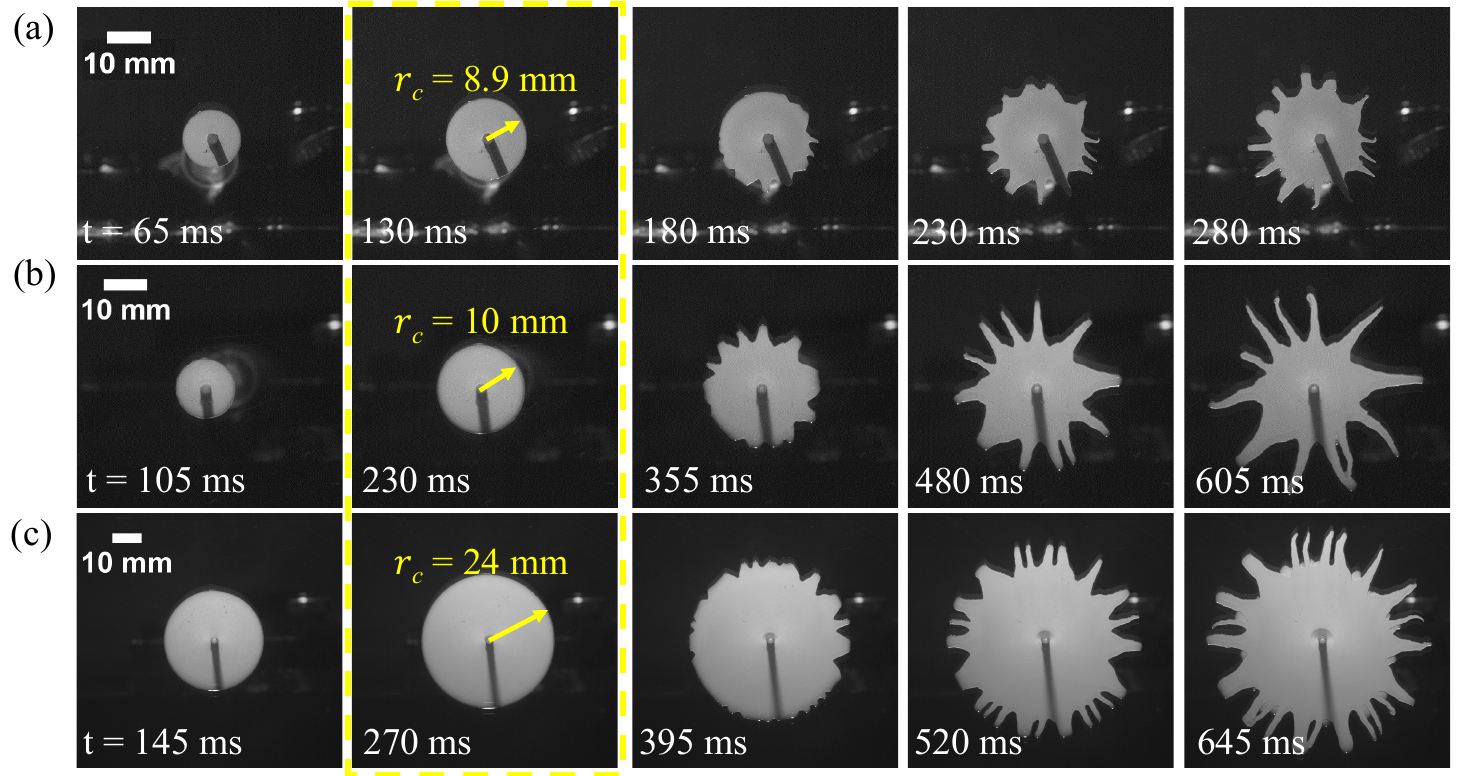}
    \caption{Snapshots for the onset of fingering instability with various flow rate $Q$ and gap thickness $h$ at $T_w = 25^{\circ}$C. (a) $Q=0.5$ mL$/$s, $h = 0.253$ mm; (b) $Q = 0.5$ mL$/$s, $h = 0.368$ mm; (c) $Q = 2$ mL$/$s, $h =0.340$ mm.}
    \label{fig:variousQ}
\end{figure}

However, at a lower $T_w =$ 25 $^{\circ}$C ($T_w <  T_m$)  (Fig. \ref{fig:observedinstability}c), the injected viscous liquid accordingly undergoes the supercooling process, and an interfacial front instability is identified with the pronounced fingering patterns along the radial direction (Supplementary Movie 1). As shown by the spatiotemporal snapshots, initially the liquid film spreads outward uniformly ($t = 305, 430$ ms), and the propagation front forms a nearly perfect circular interface. Once its radius is beyond a critical value ($r > r_c $), some little notches appear at the moving front and signify the onset of the fingering instability ($t = 555$ ms). As the fluid is continuously injected, the front progressively propagates radially, and fingers further extend longer outward ($t = 680, 805$ ms). This fingering instability is quantitatively characterized by temporal evolution of the azimuthal undulations along the interface [$r(\theta)$] as shown by Fig. \ref{fig:observedinstability}d. Note that the notches at the interface are pinning points during the subsequent growing process, and the wavelength $\lambda$ between the nearest pinning points is associated with the circumferential length and the number of the observed fingers ($N_f$), $\lambda_{exp} \sim 2\pi r_c/N_f$ (Supplementary Note 2).   

\section{Onset of the fingering instability} 
To explore the effect of flow rate ($Q$) and gap thickness ($h$) on the fingering instability, a series of experiments are performed in the Hele-shaw cell with various $Q$ and $h$ (Fig. \ref{fig:variousQ}). With the increase of gap thickness ($h$) at a fixed flow rate ($Q$), the critical time ($t_c$) for the onset of fingering instability tends to increase (Fig. \ref{fig:variousQ}a,b). With the increase of flow rate ($Q$) at a similar gap thickness ($h$), the critical radius ($r_c$) for the onset of fingering instability tends to increase (Fig. \ref{fig:variousQ}b,c).

\begin{figure}[t]
		\centering
		\includegraphics[width=0.7\textwidth]{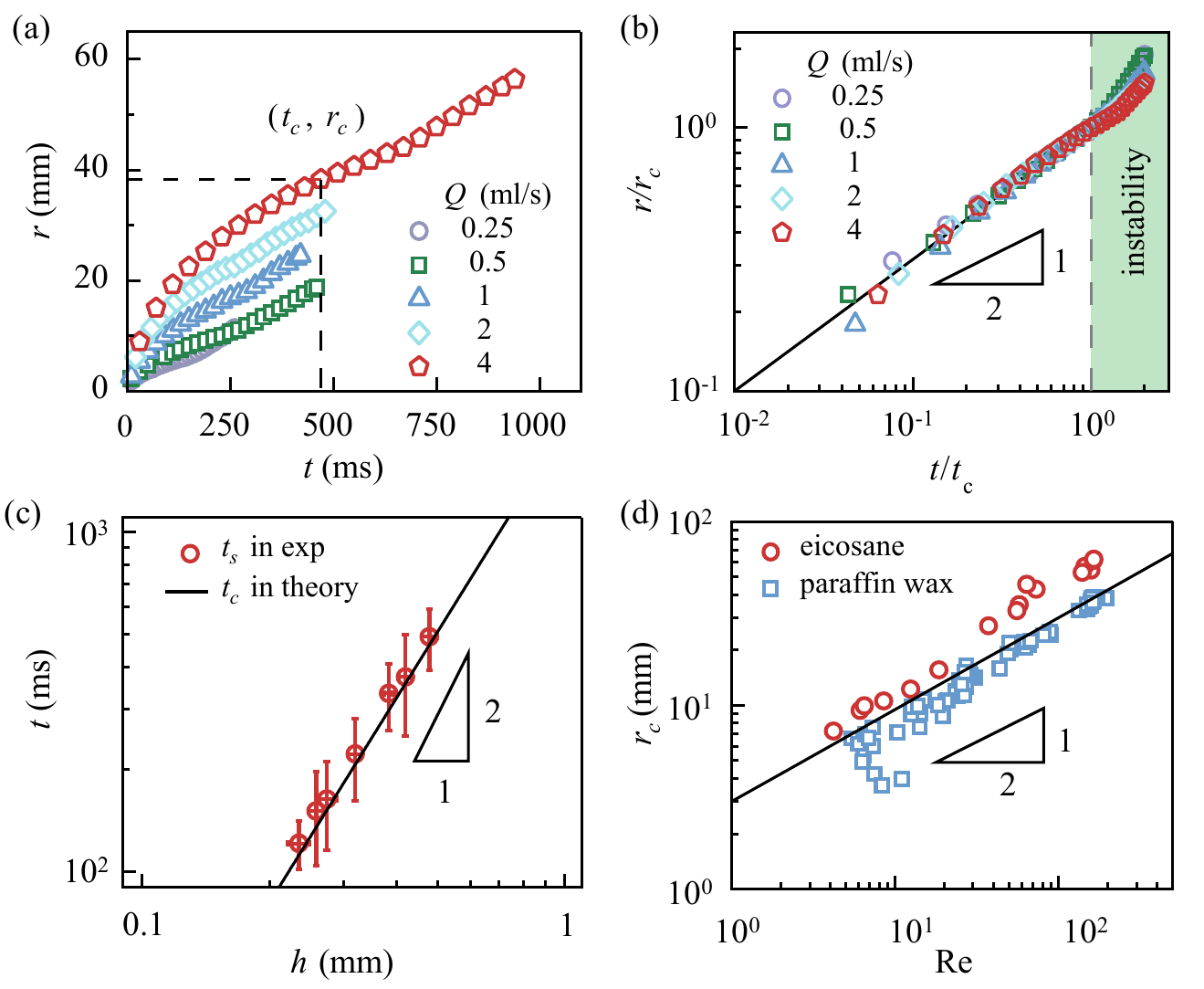}
		\caption{Scaling analysis for the onset of instability. (a) The temporal evolution of $r$  dependent on flow rates  $Q$; (b) Prior to the instability onset, the spreading radius has a power law of $1/2$ with time, $ r/ r_c \sim (t/t_c)^{1/2}$;  (c) At the onset of the instability, the experimental data of $ t_{c}$ agreeing with theory of Eq. \eqref{eq:timescalesolidification} and \eqref{eq:timescaletcqualts} excellently $t_c = t_s \sim h^2$; (d) Critical radius $ r_{c} \sim (Re)^{1/2}$. }
		\label{fig:dyanmics}
	\end{figure}

A quantitative understanding of the onset of the instability is obtained by checking the critical radius ($ r_c $) with various physical parameters (such as $Q, h$) (Supplementary Movie 2). The temporal evolution of radius ($r$) for various flow rates ($Q$) (Fig. \ref{fig:dyanmics}a) indicates the turning point for the onset of instability (a critical time $t_c$ or radius $r_c$), after which the fingering instability further develops (the radius is denoted for the length of typical fingers $r_f$). Prior to the instability, based on the mass conservation, $Q=2 \pi r h dr/dt$, the radius dependent on time is obtained,
\begin{equation}\label{eq:radiustime}
r \sim (Qt/\pi h)^{1/2} \sim t^{1/2}.
\end{equation}
As shown in Fig. \ref{fig:dyanmics}b, the experimental data collapse into a master curve with a $1/2$ scaling law, $r/r_c \sim (t/t_c)^{1/2}$.

Because of the geometric confinement in the Hele-Shaw cell, the injected viscous liquid at a high temperature $T_l$ is subjected to conductive cooling due to the bottom and top plate at a low temperature ($T_w$) at the vertical direction, likely resulting in the aforementioned instability.  By balancing the diffusive heat flux with the solidification rate \cite{ghabache2016frozen,hu2020frozen}, the timescale of solidification process at the vertical direction is obtained,
\begin{equation}\label{eq:timescalesolidification}
 t_{s}={\rho \mathcal{L}
(h/2)^{2}}/{2 \kappa \Delta T}\sim h^2,
\end{equation}
where $ \Delta T=T_m-T_w $, $\mathcal{L}$, $\rho$, $\kappa$ for the supercooling, latent heat, density and  thermal conductivity (Table I for the materials parameters). Indeed, the experimental data of $t_c$ for the onset of the instability agrees excellently with the $t_s$ for the solidification timescale at the vertical direction in theory (Fig. \ref{fig:dyanmics}c),
\begin{equation}\label{eq:timescaletcqualts}
 t_c=t_s.
\end{equation}
Additionally, the scaling law holds for various gap thicknesses as well ($t_c = t_s \sim h^2$).

Further, from Eq. \eqref{eq:radiustime}-\eqref{eq:timescaletcqualts}, the critical radius $r_c$ is attained,
\begin{equation}
  r_{c} \sim (Qh)^{1/2}(\rho \mathcal{L}/8\pi \kappa \Delta T)^{1/2} \sim Re^{1/2},
\label{eq:criticalradiusrc}
\end{equation}
where $Re=\rho v_{in}h/\mu$ ($v_{in}= Q/\pi r_{in}^2$, $\mu$ for viscosity) for Reynolds number. This $r_c\sim Re^{1/2}$ scaling is consistent with the experimental observation for paraffin wax (Fig. \ref{fig:dyanmics}d).  
                \label{fig:variousliq}

For other materials with various physical and thermal properties (Supplementary Note 3 and Table I), the $1/2$ scaling law still holds (Fig. \ref{fig:observedinstability}d) in eicosane ($\rm C_{20}H_{42}$)  with similar thermal properties of paraffin wax. But for the different thermal properties, such as liquid metal of Bi-Sn alloy with a larger thermal conductivity and smaller latent heat, $r_c$ becomes smaller to be comparable with $r_{in}$, hence the instability appears immediately right after the injection of this liquid metal.

\section{Geometry with a nonuniform gap} 
\begin{figure}[t]
		\includegraphics[width=0.9\textwidth]{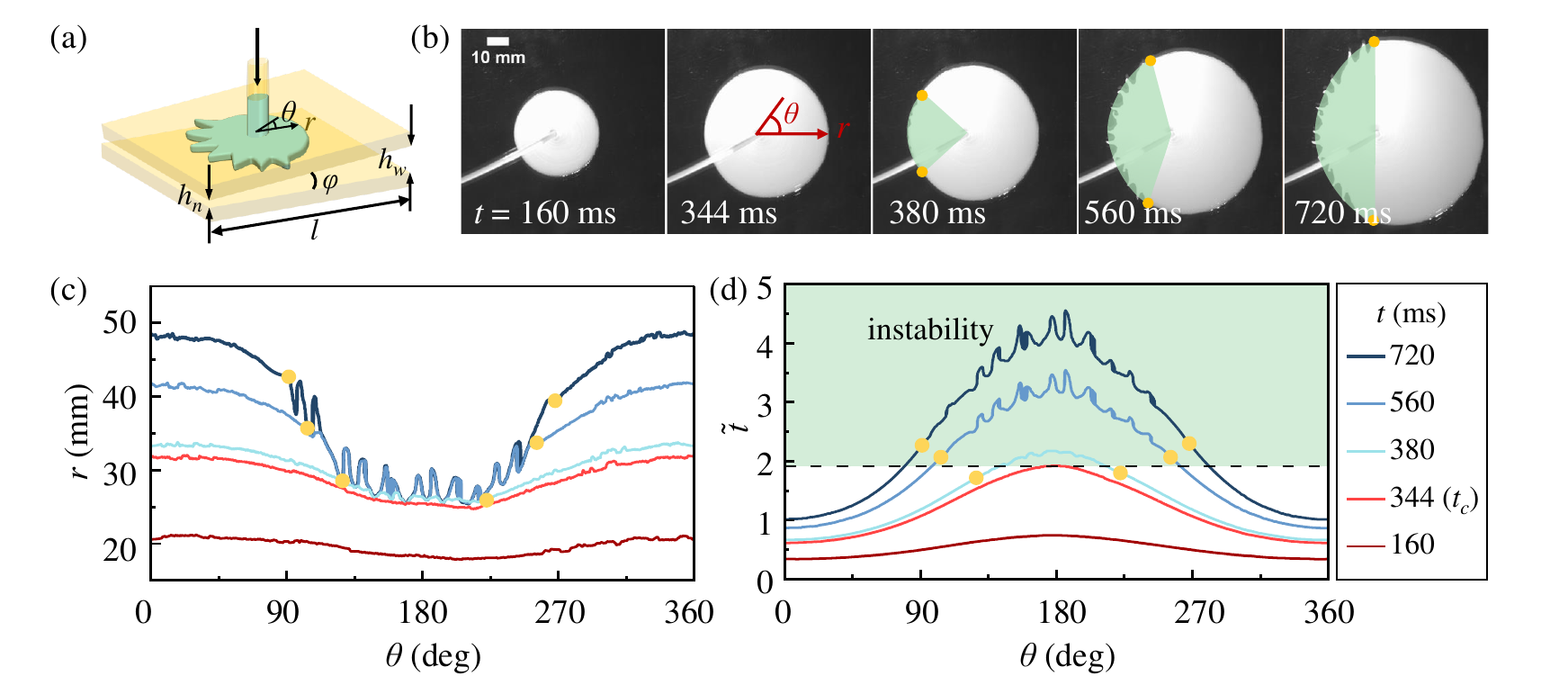}
		\caption{Geometry effect on the fingering instability. (a) Schematic diagram of the non-uniform gap thickness with a small constant gradient ($\varphi \ll 1$); (b) Interface susceptible to instability in the narrower side, while maintaining stability in the wider side; (c) $r$  at the interface as a function of azimuth at different times; (d) Emergence of fingering instability, once $\tilde{t} >2$ (as shaded in the green), consistent with the experimental observation in (b). Orange dots in (b-d) marking the region of interfacial instability.}
		\label{fig:wedgeshape}
	\end{figure}

Since the timescale of heat transfer along the vertical direction determines the onset of instability [$t_c \sim h^2$ in Eq. \eqref{eq:timescalesolidification}], we further directly demonstrate the effect of thickness in a nonuniform geometry on the resultant instability. In this wedge-shaped Hele-Shaw cell (Fig. \ref{fig:wedgeshape}a) \cite{al2012control}, the constant depth gradient is extremely small on the order of $O(10^{-3})$ \cite{expnonuniformdata}.

When the molten paraffin wax is injected into this cell at $T_w =$ 25 $^\circ$C and $Q =$ 4 ml/s,  the high-speed images show the pattern evolution (Fig. \ref{fig:wedgeshape}b and Supplementary Movie 3), and the interfacial profile $r(t, \theta)$ depends on time and the azimuthal angle (Fig. \ref{fig:wedgeshape}c). Initially the spreading profile remains a nearly circular shape ($t =$ 160 ms). Afterwards, in the left-side region with a narrower gap, the interface or the propagation front is more susceptible to the instability because of rapid cooling, and fingering instability becomes pronounced  beyond a critical time ($t_c \sim 344$ ms). Alternately, in the right-side region with a wider gap, the interface still remains stable and smooth ($t =$ 560 ms).  In this way, the asymmetry pattern is formed, \emph{i.e.}, the stability is persisted at the wider gap while the instability is triggered at the narrower gap ($t =$ 720 ms).

This geometry effect on the pattern stability is further analyzed by a simple scaling analysis. By combining equation Eq. \eqref{eq:timescalesolidification} and the local height at the interface which is described by
$h(r, \theta) = h_{in} + r(t, \theta) \rm{cos}\theta \rm{sin} \varphi$, where $h_{in}= (h_n+h_w)/2$,  the local solidification timescale at the interface is obtained,
\begin{equation} \label{eq:wedgetimescale}
t_s (r, \theta) = \rho \mathcal{L} h^2 / 8 \kappa \Delta T \sim h(r, \theta)^2.
\end{equation}

We establish the stability diagram for the interface at propagation front by defining dimensionless time $\tilde{t}(\theta)= t/t_s$ (Fig. \ref{fig:wedgeshape}d). The interface is stable for $\tilde{t} < 2$, while fingering instability sets up for $\tilde{t} > 2$. The experimental data agrees reasonably well with this scaling analysis, which is built on the heat transfer at the vertical direction by addressing the solidification timescale during the supercooling process.

\section{Proposed mechanism of thermo-viscous fingering instability}
\begin{figure}[t]
		\includegraphics[width=0.8\textwidth]{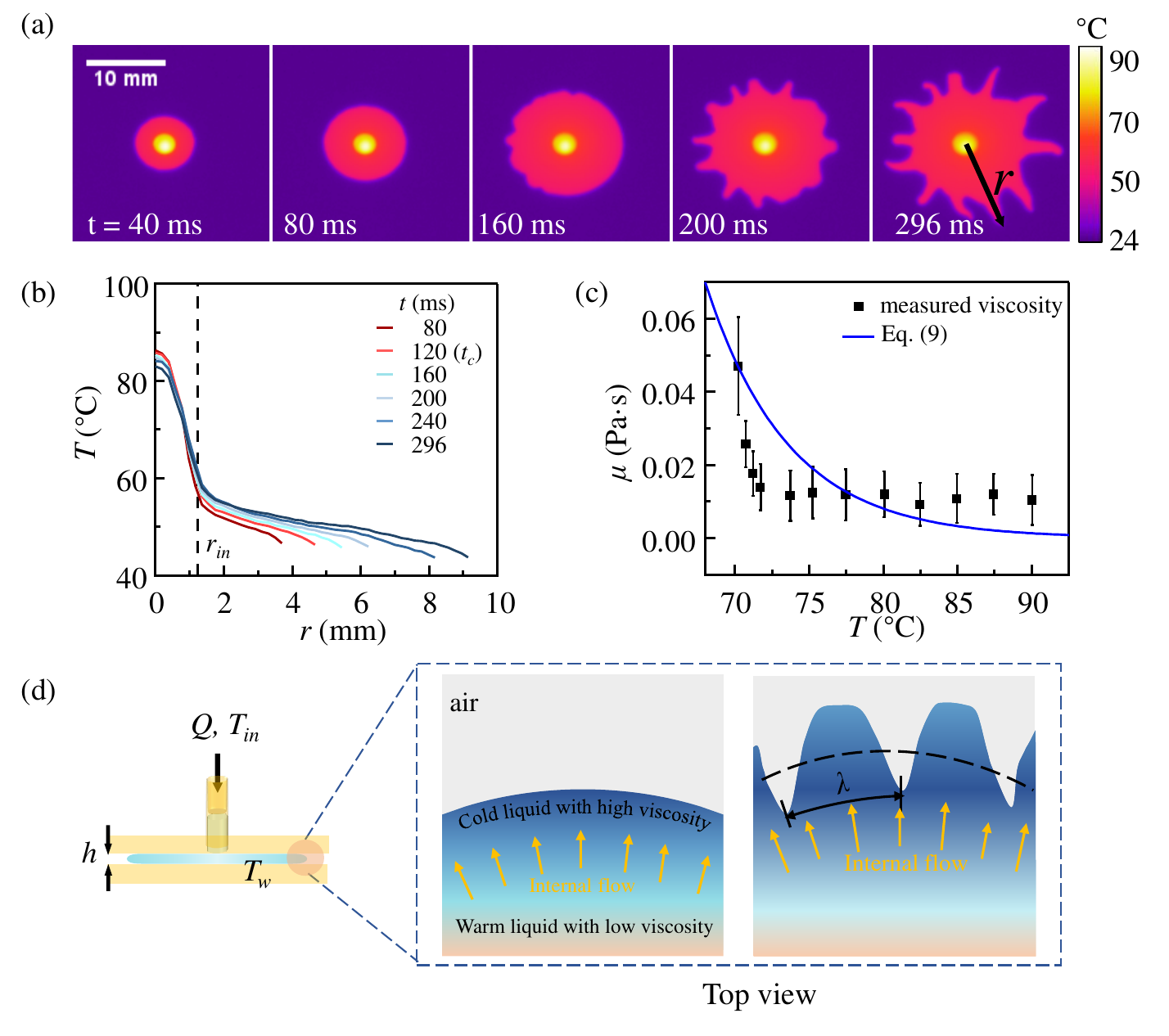}
		\caption{(a) Spatiotemporal evolution of temperature for the viscous liquid of the paraffin wax at $T_w = 24 \ ^{\circ}$C, $Q = 0.25$ ml/s, $h =$ 270 $\mu$m; (b) Temperature evolution along the radial direction of the black arrow in the last snapshot of (a); (c) The viscosity of paraffin increases significantly with the decreased temperature. (d) Sketch for the thermal-viscous fingering instability. As the molten paraffin wax passing through the thin gap, its viscosity increases sharply with the decreased temperature, and this pronounced mobility contrast between the central warm fluid and the cold fluid in the vicinity of the spreading front consequently leads to the occurrence of thermal-viscous fingering instability.}
		\label{fig:temperaturevis}
	\end{figure}

In order to unravel the role of the supercooling process in this fingering instability, the spatial-temporal evolution of  temperature for the viscous liquid is directly recorded by a thermal camera (Fig. \ref{fig:temperaturevis}a, Supplementary Movie 6, Supplementary Note 4). Fig. \ref{fig:temperaturevis}b presents temperature distribution along the radial direction of the marked finger (the black arrow in the snapshot at 296 ms). Adjacent to the inlet ($r \approx r_{in}$), the injected liquid is immediately cooled down by contacting the sidewall; then in the spreading region ($r_{in} <  r < r_f$), temperature decreases with the radius slowly. Typically paraffin wax has a strongly temperature-dependent viscosity\cite{FERRER2017154,ROSSETTI1999413,LOUANATE2021179018}, which increases rapidly with the decreasing temperature, especially when it approaches to solidification. The temperature-dependent viscosity of the used paraffin wax (Fig. \ref{fig:temperaturevis}c) is measured by a rotational rheometer (MARS3 HAAKE). As shown in the sketch of Fig. \ref{fig:temperaturevis}d, based on the observation of the decreased temperature and the increased viscosity accordingly at the spreading front, we hypothesize that the viscosity contrast between the internal warm fluid and the outer cold fluid is enhanced and the local mobility near the spreading front is much less than that in the centre, consequently leading to a thermal-viscous fingering instability \cite{Helfrich_1995,Wylie_Lister_1995}.

\section{Linear stability analysis}
\begin{figure}[t]
		\includegraphics[width=0.8\textwidth]{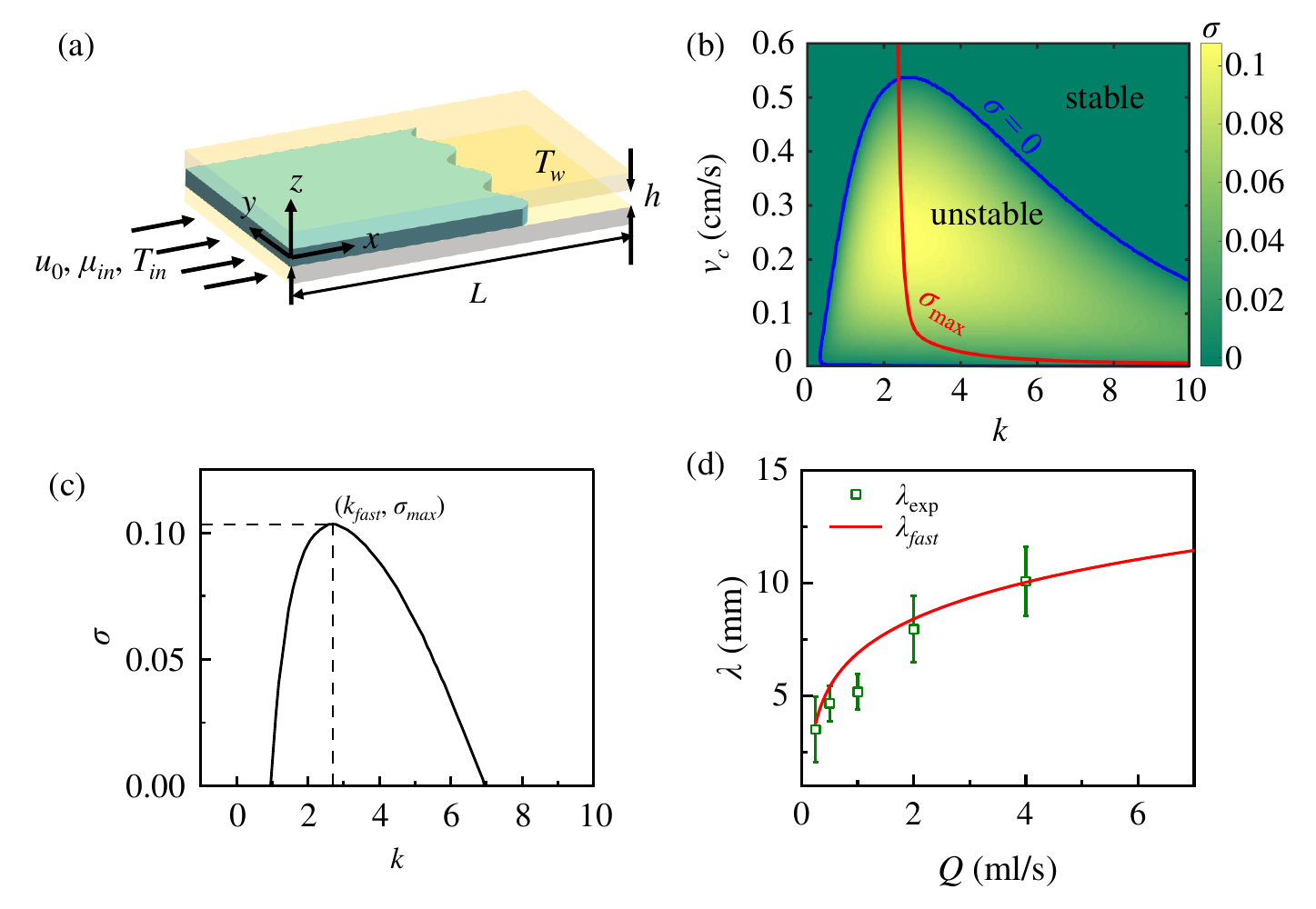}
		\caption{(a) Sketch for linear stability analysis. (b) Phase diagram of the growth rate $\sigma$, the blue and red curves for $\sigma=0$ and $\sigma_{max}$; (c) A typical dispersion relationship, and $k_{fast}$ for $\sigma_{max}$; (d) Wavelength $\lambda$ depends on injection flow rates $Q$, the line for the model, squares for experiments and error bars for five reproducible experiments.}
		\label{fig:stabilityanalysis}
\end{figure}

Despite the complicated process, the spacing between the pinning points correlated with the instability wavelength ($\lambda_{exp}$) is reproducible and increases with the flow rate $Q$ (see Fig. \ref{fig:stabilityanalysis}d). The linear analysis in a 2D radial coordinate frame will be much more complicated mathematically, such as the radial flow decays with time as the front move radially.  Here to link the previous theory in 2D (x,y) coordinate system \cite{Helfrich_1995} with the current work, we obtain the instability wavelength by employing the local spreading velocity at the propagation front ($v_f=\mathrm{d}r/\mathrm{d}t$, rather than $v_{in}$), implicitly including the radial effect in our experiments. Also, the ratio between $r_c/r_{in}\sim 10$ during the instability occurrence implies the negligible curvature effect. Specifically, the linear stability analysis is performed within a rectangle Hele-Shaw cell to seek the relation between the uniform constant inlet velocity and the instability growth. This uniform constant inlet velocity in the rectangle Hele-Shaw cell refers to the front velocity $v_f$ in the present radial flow. Then we can obtain the relation between the instability and critical velocity $v_c$, \emph{i.e.} $v_{f}(r=r_c)$ which varies with the inlet flow rate $Q$.

As sketched in Fig. \ref {fig:stabilityanalysis}a, the hot liquid at $T_{in}$ with a uniform constant inlet velocity $u$ and viscosity $\mu(T)$ is forced into the gap, where $L$ is the length of the rectangle Hele-Shaw cell. Since the flow here is in the horizontal plane and the thickness of the liquid film is much smaller than the length scale of the flow, the effects of gravity and buoyancy are ignored. In our experiments, the Peclet number is $ Pe=v_{in} r_c/\alpha \sim 10^5$ for the typical values $v_{in} \sim 1$\ m/s, $r_c\sim 0.01$ m and the thermal diffusivity of paraffin $\alpha=10^{-7}\mathrm{\ m^2/s}$, so within the flow plane we only consider the heat advection, while in the vertical direction we only consider the heat loss to the sidewall.

By simply following the linear-stability analysis in \cite{Helfrich_1995}, the governing equations for the Darcy’s law, incompressible flow velocity and heat transfer are modified accordingly as below:
\begin{equation}
\frac{12 \mu}{h^2} \mathbf{u}=-\nabla p,\\
\label{governereq1}
\end{equation}
\begin{equation}
\boldsymbol{\nabla} \cdot \mathbf{u}=0,\\
\label{governereq2}
\end{equation}
\begin{equation}
\frac{\partial T}{\partial t}+\mathbf{u} \cdot \nabla T=-\delta^{\prime}\left(T-T_w\right).
\label{governereq3}
\end{equation}
Here  $\mathbf{u}=(u, v)$ and $T$ are the gap-averaged velocity and temperature, $t$ the time,  $p$ the pressure. Heat loss to the side walls is approximated by the last term in Eq. \eqref{governereq3} with the heat transfer coefficient $\delta^{\prime} \approx \frac{\pi^2 \alpha}{h^2}$.

Generally, the viscosity for wax varies with temperature in a much more complicated way, but for the simplicity of theoretical analysis, we assume that the thermo-physical parameters of the paraffin wax are constant except for the viscosity variation to be simplified as a function of the temperature decrease with the constant coefficient $\gamma^{\prime}$ as following \cite{Helfrich_1995},
\begin{equation}
\mu=\mu_{m} \mathrm{e}^{\gamma^{\prime}\left(T_{m}-T\right)}.
\label{viscosity}
  \end{equation} 
Here the value of $\gamma^{\prime}$ is set to be 0.18 per Celsius degree \cite{FERRER2017154,LOUANATE2021179018}, $\gamma=\gamma^{\prime}(T_{m}-T_w)=7.2$ with the temperature difference $T_{m}-T_w\approx 40\ ^{\circ}$C. In addition, the measured temperature-dependent viscosity is comparable with the simplified model Eq. \eqref{viscosity} (Fig. \ref{fig:temperaturevis}c).
    
The governing equations and the viscosity Eq. \eqref{viscosity} are then normalized with nondimensional parameters $(\tilde{x}, \tilde{y})=(x,y)/L$, $\tilde{t}=tU/L$, $(\tilde{u}, \tilde{v})=(u, v)/(Ph^2/12L\mu_{m}^2)$, $\tilde{\theta}=(T-T_w)/(T_{m}-T_w)$, $\delta=\delta^{\prime}L/U$, $\gamma=\gamma^{\prime}(T_{m}-T_w)$. The velocity scale $U=Ph^2/12L\mu_{m}^2$ $U$ is obtained from the Poiseuille relation for fluid with viscosity $\mu_{m}$ and the driven pressure gradient of $P/L$.

By eliminating $p$ with the stream function defined as $(\tilde{u}, \tilde{v})=(\psi_y,-\psi_x)$, Eq. \eqref{governereq1} gives,
\begin{equation}
	\nabla^2\psi=\gamma\nabla\psi\cdot\nabla\tilde{\theta}.
	\label{vorticityequ}
\end{equation}
Then Eq. \eqref{governereq3} become as below, respectively,
\begin{equation}
\frac{\partial \tilde{\theta}}{\partial t}+J(\psi, \tilde{\theta})=-\delta \tilde{\theta},
	\label{temperatureequ}
\end{equation}
with the Jacobian $J(a, b)=a_x b_y-a_y b_x$.

The inlet velocity and temperature are kept at $\tilde{u}_0$ and $T_{in}$, so the inlet boundary conditions are:
\begin{equation}
	\tilde{\theta}=1 \quad at \quad \tilde{x}=0,
	\label{inletBCfortheta}
\end{equation}
\begin{equation}
	\tilde{u}=\tilde{u}_0 \quad at \quad \tilde{x}=0.
	\label{inletBCforpsi}
\end{equation}
With the assumption of uniform oulet pressure for Eq. \eqref{governereq1}, the outlet boundary condition is:
\begin{equation}
	\tilde{v} =-\psi_x=0 \quad at \quad \tilde{x}=1.
	\label{outletBCforpsi}
\end{equation}
By neglecting the variations and flow in the $y$ direction, a steady basic state with the constant inlet $x$-velocity $\tilde{u}_0$ is:
\begin{equation}
	\Psi=\tilde{u}_0\tilde{y},\quad \Theta=\mathrm{e}^{-\frac{\delta \tilde{x}}{\tilde{u}_0}},\quad \mathcal{M} =\mathrm{e}^{\gamma(1-\Theta)}.
	\label{basicstate}
\end{equation}

Here $\Psi$, $\Theta$, $\mathcal{M}$ are the basic-state stream function, temperature and viscosity, respectively. The pressure drop between $\tilde{x}=0$ and $1$ is
\begin{equation}
  	\Delta p=1=\tilde{u}_0\int_0^1 \mathcal{M}(\tilde{x}) \mathrm{d}\tilde{x}.
  	\label{intpressure}
  \end{equation}

From Eq. \eqref{intpressure}, for a certain value of $\gamma$, the relation between the inlet velocity $\tilde{u}_0$ and heat transfer coefficient $\delta$ can be numerically solved. By employing the perturbations $\phi=f(x)\text{sin}(ky)\mathrm{e}^{\sigma t}$ and $\tilde{\theta}^{\prime}=g(x)\text{cos}(ky)\mathrm{e}^{\sigma t}$ with the wavenumber $k$ in the $y$-direction and the growth rate $\sigma$ from the basic-state $\Psi$, $\Theta$:
\begin{equation}
	\psi=\Psi+\phi(x, y, t) ; \quad \tilde{\theta}=\Theta+\tilde{\theta}^{\prime}(x, y, t),
	\label{perturbedstate}
\end{equation}
Eq. \eqref{vorticityequ} and Eq. \eqref{temperatureequ} become as below,
\begin{equation}
	\nabla^2 \phi=\gamma\left(\tilde{u}_0+\phi_y\right)\tilde{\theta}_y^{\prime}+\gamma\left(\Theta_x+\tilde{\theta}_x^{\prime}\right) \phi_x,
	\label{purterbationeq1}
\end{equation}
\begin{equation}
	\tilde{\theta}_t^{\prime}+\tilde{u}_0 \tilde{\theta}_x^{\prime}+\phi_y \Theta_x+J\left(\phi, \tilde{\theta}^{\prime}\right)=-\delta \tilde{\theta}^{\prime}.
	\label{purterbationeq2}
\end{equation}
The boundary condition on $\tilde{\theta}^{\prime}$ (from Eq. \eqref{inletBCfortheta}, \eqref{basicstate} and \eqref{perturbedstate}) and $\phi$ (for inlet from Eq. \eqref{inletBCforpsi}, \eqref{basicstate} and \eqref{perturbedstate} while for outlet from Eq. \eqref{outletBCforpsi}) are,
\begin{equation}
	\tilde{\theta}^{\prime}=0 \quad at \quad \tilde{x}=0,
	\label{inletBCforperturedtheta}
\end{equation}
\begin{equation}
	\phi_y=0 \quad at \quad \tilde{x}=0,
	\label{inletBCforperturedphi}
\end{equation}
\begin{equation}
	\phi_x=0 \quad at \quad \tilde{x}=1.
	\label{outletBCforperturedphi}
\end{equation}

From Eq. \eqref{purterbationeq1} and Eq. \eqref{purterbationeq2} and the boundary conditions Eq.\eqref{inletBCforperturedtheta}, \eqref{inletBCforperturedphi} and \eqref{outletBCforperturedphi}, the eigenvalue problem for the growth rate $\sigma$ and structure functions $f(x)$ and $g(x)$ is obtained,
\begin{equation}
\frac{\mathrm{d}^2 f}{\mathrm{d}x^2}-\gamma \frac{\mathrm{d} \Theta}{\mathrm{d} x} \frac{\mathrm{d} f}{\mathrm{~d} x}-k^2 f+k\gamma \tilde{u}_0g=0, \\
\end{equation}

\begin{equation}
\tilde{u}_0\frac{\mathrm{d} g}{\mathrm{d} x}+(\sigma+\delta) g+k \frac{\mathrm{d} \Theta}{\mathrm{d}x} f=0, \\
\end{equation}
with
\begin{equation}
g(0)=0, \quad \frac{\mathrm{d} f(1)}{\mathrm{d} x}=0, \quad f(0)=0.
\end{equation}

Using the Runge-Kutta method with shooting in MATLAB, this eigenvalue problem can be solved numerically to find the growth rate $\sigma$ dependent on $\tilde{u}_0$ and the wavenumber $k$. For the rectangle Hele-Shaw cell, the dimensionless velocity $u$ can be obtained by typically taking the characteristic velocity as $U=Ph^2/12L\mu_{m}^2$. However, in the radial Hele-Shaw flow, it is hard to find a characteristic velocity since the flow decelerates during the spreading process. Here we simply use a constant velocity $v_p$ to obtain the dimensional critical unstable velocity $v_c=v_p\tilde{u}_0$.

As shown in Fig. \ref{fig:stabilityanalysis}b, the flow is unstable only when the front velocity slows down, which is consistent with the experimental observation (Supplementary Note 5). The most unstable wavenumber $k_{fast}$ for $\sigma_{max}$ increases with the decreased critical unstable velocity $v_c$. As for a typical critical unstable velocity, the dispersion relationship $\sigma(k)$ is shown in Fig. \ref{fig:stabilityanalysis}c, which indicates that the instability has both long- and short-wave cutoffs.

The relation between the critical unstable velocity $v_c$ and the inlet flow rate $Q$ is attained from the mass conservation $Q=2 \pi r h v_f$  together with Eq. \eqref{eq:criticalradiusrc},
\begin{equation}
v_c=\frac{Q}{2\pi r_c h}\sim \frac{Q}{2\pi (Qh)^{1/2} h}\sim Q^{1/2}.
\label{eq:vc_Q}
\end{equation}
From Eq. \eqref{eq:vc_Q} and the definition of the most unstable wavelength as $\lambda_{fast}=2\pi r_p/k_{fast}$, where $r_p$ is taken as the typical critical radius from the experiments, we can achieve the theoretical $\lambda_{fast}$ as a function of the flow rate $Q$, as shown in Fig. \ref{fig:stabilityanalysis}d. the observed instability wavelength is quantitatively comparable with the fastest instability growth mode. For large $k$, $v_c$ decreases to a certain value, correspondingly to the left cutoff of $\lambda_{fast}$ in Fig. \ref{fig:stabilityanalysis}d. 

\section{Discussion and outlook}
The vertical-supercooling-controlled fingering instability offers several unique advantages to further survey and control fluid instabilities in general. Firstly, the spreading of the injected jet is confined between the two plates at a moderate $Re$ number down to $Re \sim O(1, 10^2)$, circumventing the complicated dynamics in the other relevant systems, such as the droplet impacting on the substrate with splashing, retreating and bouncing \cite{yarin2006drop}. Although the radial fingering instability can be found during droplet impacting, this phenomenon originates from a completely different mechanism of Rayleigh-Taylor instability, as the interfacial front undergoes a strong deceleration at a high Reynolds number up to $Re \sim O(10^4)$ \cite{thoroddsen1998evolution}.

Secondly, here two states of the stability or instability (fingering patterns) are demonstrated by simply setting or gating the side-wall temperature at either an elevated temperature or a low temperature. At an elevated temperature, the fingering instability is inhibited; but once the gated temperature at the side wall is below the melting temperature, the instability is triggered by the thermal-viscous effect during the supercooling process. The timescale for the onset of instability is plainly determined by the solidification timescale during the supercooling process [$t_c =t_s$ of Eq. \eqref{eq:timescaletcqualts}].

Thirdly, the fluid instabilities can be engineered through the structure design in a Hele-Shaw cell. As shown in Fig. \ref{fig:wedgeshape}, the asymmetry pattern is obtained in the wedged geometry, and more sophisticated structures together with the patterned surfaces \cite{kataoka1999patterning} are expected to extend our capability to further engineer the liquid flow.

Lastly, despite the simplified model, the key role of viscosity contrast during the solidification process has been essentially captured in the theory, evidently highlighting the underlying physical mechanism for the observed fingering instability here. A more accurate and complete theory is beyond the scope of this work and will be worthy of thorough investigation in the future. Nevertheless, considering the simplified model for such complicated coupling processes of hydrodynamics, heat transfer and phase change, this remarkable agreement between theory and experiment clearly validates the proposed mechanism based on the thermo-viscous fingering instability during the solidification, which is determined by the ambient cooling at the vertical direction.


This work presents that the strategy of the vertical supercooling and the structure design to control and steer the liquid flow and the subsequently correlated instabilities for the potential technological implications, such as microfluidic cooling for semiconductor devices \cite{microfluidicscoolingnature}. Particularly, this supercooling process of the molten jet and liquid film is inherent within the state-of-the-art thermal drawing process, which involves heating and the subsequent cooling stage, and consequently sheds light on the diverse fluid instabilities to fabricate versatile functional structures for the advanced functional fibres and fabrics \cite{AbouraddyNatureMat2007,KaufmanNature2012,DengAFM2020,DengNatureCom2022}.

\begin{acknowledgments} D. D. is grateful to Prof. Howard Stone, Prof. K. L. Chong and Prof. Martin Bazant for the insightful and supportive comment on this work. We would like to acknowledge the editor and the anonymous referees for the constructive suggestions to improve this manuscript as well, and Y. X. Qin for the viscosity measurement. This work is supported by the funding from the National Program in China and startup in Fudan University.
\end{acknowledgments}

	\bibliographystyle{ScienceAdvances.bst}

\end{document}